\documentclass[conference]{IEEEtran}
\IEEEoverridecommandlockouts
\usepackage{cite}
\usepackage{amsmath,amssymb,amsfonts}
\usepackage{algorithmic}
\usepackage{graphicx}
\usepackage{multicol}
\usepackage{multirow}
\usepackage{tikz}
\usepackage{makecell}
\usepackage{textcomp}
\usepackage{xcolor}
\usepackage{subcaption}
\newcommand\copyrighttext{%
  \footnotesize \textcopyright 2022 IEEE. Personal use of this material is permitted. Permission from IEEE must be obtained for all other uses, in any current or future media, including reprinting/republishing this material for advertising or promotional purposes, creating new collective works, for resale or redistribution to servers or lists, or reuse of any copyrighted component of this work in other works.
}
\newcommand\copyrightnotice{%
\begin{tikzpicture}[remember picture,overlay]
\node[anchor=south,yshift=10pt] at (current page.south) {\fbox{\parbox{\dimexpr\textwidth-\fboxsep-\fboxrule\relax}{\copyrighttext}}};
\end{tikzpicture}%
}
  
\newcommand{\model}{SpecRNet}
\def\BibTeX{{\rm B\kern-.05em{\sc i\kern-.025em b}\kern-.08em
    T\kern-.1667em\lower.7ex\hbox{E}\kern-.125emX}}
\begin{document}

\title{SpecRNet: Towards Faster and More Accessible Audio DeepFake Detection\\
{}
\thanks{This work is partially supported by Polish National
Science Centre --- project UMO-2018/29/B/ST6/02969.}
}

\author{
\IEEEauthorblockN{Piotr Kawa, Marcin Plata and Piotr Syga}
\IEEEauthorblockA{\textit{Department of Artificial Intelligence} \\
 \textit{Wrocław University of Science and Technology} \\
Wrocław, Poland \\
\textit{\{piotr.kawa,marcin.plata,piotr.syga\}@pwr.edu.pl}}
}

\maketitle
\copyrightnotice

\begin{abstract}
Audio DeepFakes are utterances generated with the use of deep neural networks.
They are highly misleading and pose a threat due to use in fake news, impersonation, or extortion. 
In this work, we focus on increasing accessibility to the audio DeepFake detection methods by providing \model{}, a neural network architecture characterized by a quick inference time and low computational requirements. Our benchmark shows that \model{}, requiring up to about 40\% less time to process an audio sample, provides performance comparable to LCNN architecture --- one of the best audio DeepFake detection models. Such a method can not only be used by online multimedia services to verify a large bulk of content uploaded daily but also, thanks to its low requirements, by average citizens to evaluate materials on their devices.
In addition, we provide benchmarks in three unique settings that confirm the correctness of our model. They reflect scenarios of low--resource datasets, detection on short utterances and limited attacks benchmark in which we take a closer look at the influence of particular attacks on given architectures.
\end{abstract}


\section{\uppercase{Introduction}}
\label{sec:introduction}

DeepFakes refer to a branch of algorithms for manipulating audio--visual content. They utilize deep neural networks to create highly convincing spoofs of biometric features such as the face or voice.
It is one of the most popular methods of automatic generation or manipulation of audio--visual modifications of the original content. 
The original approach was introduced in 2017 and allowed to create a video in which the face of the original individual was swapped with the face of another person. 
This phenomenon is also present in the field of audio. The term ''Audio DeepFake'' covers solutions that can create artificially modified speech. These solutions can either generate new utterances using Text--To--Speech (TTS)~\cite{tacotron,tacotron-2,mellotron} and Voice Cloning~\cite{rtvc,vc-neurips} methods, or modify existing utterances and therefore change it to someone else --- Voice Conversion~\cite{stargan_first,sc-glowtts}.
The more recent architectures not only provide a synthesis of a speech but also focus on proper intonation, stress, and rhythm.
Such tampered material is characterized by high quality and is highly misleading. DeepFake utterances often achieve high mean opinion scores compared to bona fide utterances.

DeepFakes have many malicious applications and can hinder various spheres of life. Fake news is one the most prominent and dangerous examples --- e.g. induction of the conflicts by depicting politicians saying bogus things.
Tampered samples can also be used to bypass speaker recognition (authorization) systems (e.g. voice--recognition in banking systems) which are increasingly popular due to directives like~\cite{psd2}. Furthermore, DeepFake--based impersonation is also used for extortion~\cite{audio-scam}.

The process of creating DeepFake audio is simple --- there exist many well documented open--source toolkits that require only a consumer--grade computer. In addition, methods such as Real Time Voice Cloning (RTVC)~\cite{rtvc} generate voice cloning utterances using only a few seconds of voice signal. The results are less sophisticated than DeepFakes created with state--of--the--art methods using hours of the source material. Nonetheless, they can still be used to deceive people --- especially when combined with low--quality Internet connection and other factors that decrease the audio quality and therefore make manipulations even more difficult to spot.

Audio spoofing is a problem from the field of speech processing which is similar to the concept of DeepFakes. It covers problems of voice synthesis, voice conversion, or replay attacks. The key difference between DeepFakes and spoofing is the target one method wishes to deceive. While DeepFakes are mainly used to trick humans so they believe utterance is authentic, spoofing methods aim to deceive automatic speech verification systems that a person being verified is an owner of a particular voice. In addition, DeepFakes always generate new material, whereas audio spoofing can be created from existing samples, e.g. via replaying or merging various audio chunks.

Constantly raising threat of DeepFakes caused by more and more advanced methods, their increasing popularity and ease of creation, induced the scientific community to research methods of determining whether a given utterance is pristine or generated. The introduced solutions vary both in terms of the classification methods (e.g. Gaussian Mixture Models (GMMs)~\cite{gmm} or deep neural networks~\cite{lcnn,selcnn,fakeavceleb,rawnet2}) as well as the representation of the audio they base on. 

Our primary motivation is to decrease the computational requirements and inference time of audio DF detection. For this purpose, we introduce \model{} --- a novel spectrogram--based model inspired by RawNet2~\cite{rawnet2} backbone. Less demanding architectures can have a wide range of applications allowing quick screening of the suspicious samples.

The problem of DeepFakes should be considered at scale --- the volume of the potentially considered materials can qualify it as a big data task.
New legal regulations are introduced (e.g. European Union's Strengthened Code of Practice on Disinformation~\cite{eu_deepfakes}) that obligate content providers and hosts to implement self--regulatory standards to counteract against disinformation, including DeepFakes. 
This means that further steps in research of DeepFake detection area should aim to decrease their computational complexity while providing high performance. 
The most natural application of those methods is in online multimedia platforms, where 720,000 hours of new videos is added daily~\cite{yt-stats}. Such platforms are often used for spreading fake news, and lightweight detection methods can be handy to spot potentially forged materials. 

In addition, such lightweight methods give a possibility to an average citizen to detect forged materials --- this is achieved by reducing the computational and memory requirements to run the detection method. Note that many such citizens may not be able to use GPUs in order to execute the detection.
We point out that due to the small quantity of the source material the average citizen will likely be targeted with a less sophisticated (hence less convincing) forgeries. This means that smaller models, despite not achieving state--of--the--art results are still capable of detecting such manipulations.
Such topics were previously raised in the following works~\cite{mesonet,verify-it-yourself, defakehop}.

Our contribution in this paper includes:
\begin{itemize}
    \item proposing a new neural network architecture --- \model{},
    \item comparing the performance of \model{}, to its inspiration --- RawNet2~\cite{rawnet2} and a one of the leading architectures for Audio DeepFake Detection -- LCNN~\cite{lcnn} on WaveFake dataset,
    \item time evaluation on both CPU and GPU with respect to different batch sizes,
    \item evaluation on three unique settings --- data scarcity, limited attacks and short utterances.
\end{itemize}

\section{Related work}

Detection of audio DeepFakes is a more recent task than its visual equivalent. For this reason, the number of both audio datasets and detection methods is much smaller.
ASVspoof~\cite{asvspoof2021} is one of the most important challenges regarding the detection of spoofed audio. The 2021 version of this bi--yearly challenge brought a new subset --- along with logical access (LA) and physical access (PA), there exists now a speech DeepFake (DF) subset. This new subset was not yet available when this paper was written.
FakeAVCeleb~\cite{fakeavceleb} is an example of multi--modal datasets --- the samples are composed of both visual and audio manipulations. While the visual part of the dataset was composed of many methods, its audio equivalent was generated using only one approach --- RTVC~\cite{rtvc}. Thus, it was not selected by us as the evaluation subset due to the small variety of generated samples in relation to other datasets.
WaveFake~\cite{wavefake} is, up to this day, the largest audio DeepFake detection dataset. It consists of about 120,000 samples, and its generated samples were created using 8 spoofing methods. Due to its volume, a number of supported languages (English and Japanese), and generation methods, it became our choice as a benchmark data.

Audio DeepFake detection, despite lesser renown than its visual equivalent, is ever--growing branch of validity verification. Due to its resemblance to audio spoofing, some of the spoofing detection methods were adapted to this field as well.
Detection approaches can be divided into classical and deep--learning--based methods. Gaussian Mixture Models (GMM)~\cite{gmm} is one of the most prominent examples of the classical approaches. However, nowadays, the majority of the methods are based on deep learning (DL). One of the reasons behind that is that classical approaches like GMMs tend to require a separate model for each attack type, which decreases the scalability and flexibility of such solutions. DL--based methods are further divided on the basis of the used audio representation. Methods like~\cite{rawnet2,rawgat-st} process a raw signal information. Solutions like~\cite{lcnn,selcnn,attack-agnostic-dataset} base on a spectrogram representation of audio form (front--ends). Mel--frequency cepstral coefficients (MFCCs) and linear--frequency cepstral coefficients (LFCCs) are one of the most popular approaches. In addition, methods of visual DeepFake detection, based on spectrogram features, were lately adapted to audio DeepFake detection~\cite{fakeavceleb}.

\section{\model{}}

In this section, we present a novel architecture --- \model{}. Its backbone is inspired by RawNet2~\cite{rawnet2}. Contrary to its predecessor, which operates on one--dimensional raw audio signal, it processes two--dimensional spectrogram information --- in particular linear frequency cepstral coefficients (LFCC). This decision was inspired by the recent works that show a significant increase in performance of the spectrogram--based models in relation to architectures based on raw signals~\cite{wavefake}.

\begin{table}[htb]
\caption{The scheme of SpecRNet architecture. The convolution layers are defined according to the convention: \textit{Conv2D(kernel size, input channels, output channels)}, the layers under dotted lines in residual blocks refer to operations performed on identity tensors before adding.}\centering
\label{tab:architecture}
\begin{tabular}{|c|c|c|}
    \hline
    \textbf{Layer} & \textbf{Input} & \textbf{Output shape}  \\ \hline 
    \hline
    LFCC & 64600 & $ 1 \times 80 \times N$ \\ \hline
    \makecell{preliminary \\ normalization} &  \makecell{BN2D \\ SELU} & $1 \times 80 \times N$ \\ \hline
     ResBlock & \makecell{Conv2D(3, 1, 20) \\ BN2D \\ LeakyReLU(0.3) \\ Conv2D(3, 20, 20) \\ $\cdots \cdots \cdots \cdots \cdots$ \\ Conv2D(1, 1, 20)} & $20 \times 80 \times N$ \\ \hline
     FMS Attention & \makecell{Maxpool(2) \\ FMS \\ Maxpool(2)} & $20 \times 20 \times \frac{N}{4}$ \\ \hline
     ResBlock & \makecell{BN2D \\ LeakyReLU(0.3) \\ Conv2D(3, 20, 64) \\ BN2D \\ LeakyReLU(0.3) \\ Conv2D(3, 64, 64) \\ $\cdots \cdots \cdots \cdots \cdots$ \\ Conv2D(1, 20, 64)} & $64 \times 20 \times \frac{N}{4}$ \\ \hline
     FMS Attention & \makecell{Maxpool(2) \\ FMS \\ Maxpool(2)} & $64 \times 5 \times \frac{N}{16}$ \\ \hline
     ResBlock & \makecell{BN2D \\ LeakyReLU(0.3) \\ Conv2D(3, 64, 64) \\ BN2D \\ LeakyReLU(0.3) \\ Conv2D(3, 64, 64) \\ $\cdots \cdots \cdots \cdots \cdots$ \\ - } & $64 \times 5 \times \frac{N}{16}$ \\ \hline
     FMS Attention & \makecell{Maxpool(2) \\ FMS \\ Maxpool(2)} & $64 \times 1 \times \frac{N}{64}$ \\ \hline
	\makecell{pre-recurrent \\ normalization} &  \makecell{BN2D \\ SELU} & $64 \times 1 \times \frac{N}{64}$  \\ \hline
	GRU & GRU(64, bi) & 128 \\ \hline
	GRU & GRU(128, bi) & 128 \\ \hline
    FC & 128 & 128  \\  \hline
	FC & 128 & 1 \\ \hline
\end{tabular}
\end{table}

\begin{figure}[h!]
    \begin{center}
      \includegraphics[width=0.55\linewidth]{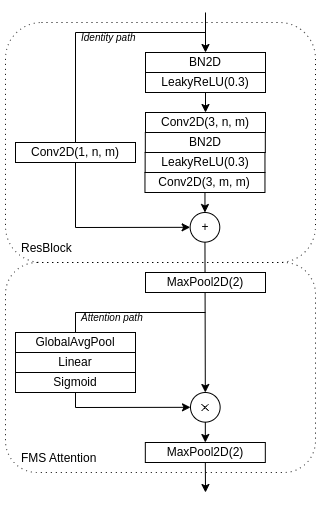}
    \end{center}
    \caption{The visualization of the residual block (ResBlock) and the FMS attention block.}
    \label{fig:res_fms}
\end{figure}

\model{} takes as an input LFCC representation of an audio signal. Network's input is first normalized using two--dimensional batch normalization~\cite{batch-normalization} followed by the SeLU activation function. 
Model is later composed of 3 residual blocks containing 2 convolutional layers preceded by batch normalization layers and LeakyReLU activation functions. Only the first residual block skips the normalization and activation layers at the beginning as the input is being processed by the batch normalization and SeLU layers. We also utilize an additional convolution layer with a kernel size of 1 on the residual identity path. The convolution layer is used to synchronize the number of channels between the identity and main paths. The blocks, similarly to RawNet2, come in two kinds differing in the number of convolution channels. Note that, due to the constant number of channels (64) in the third residual block, no synchronization between the identity and main paths is required, thus the additional convolution layer is not applied in this case. Next, each block is succeeded by two--dimensional max pooling layer, FMS attention block~\cite{rawnet2_original} and the next two--dimensional max pooling layer. The forward pass through the residual block and the FMS attention layer is presented in Figure~\ref{fig:res_fms}. 

The scheme of residual blocks does not differ from the original apart from operating on 2D thereby helping to emphasize the most informative filter outputs. The architecture ends with two bidirectional GRU layers~\cite{gru} followed by two fully connected layers. The model returns a single value in a range $[0,1]$ that after applying a threshold indicates the input as either fake or bona fide. Detailed architectural parameters are presented in Table~\ref{tab:architecture}. We also provide PyTorch implementation which is available on GitHub:~{https://github.com/piotrkawa/specrnet}.

\section{Benchmark --- full dataset}\label{sec:full_dataset_benchmark}

In order to determine our model performance we conducted benchmark tests using WaveFake dataset~\cite{wavefake}. Our implementation was based on the codebase provided by its authors. We compare our model with two aforementioned state-of-the-art architectures, namely LCNN and  RawNet2. We selected these architectures because LCNN and RawNet2 represent two approaches to problem of DeepFake detection --- analysis of respectively spectrogram--based information and raw audio signal. These architectures are also methods typically used in the task of detecting audio spoofing.

\subsection{Dataset}

The dataset used in the benchmark is composed of 104,885 fake and 13,100 bona fide samples. Pristine utterances were collected from two datasets --- English LJSpeech~\cite{ljspeech17} and Japanese JSUT~\cite{jtsu}.
They were also used to create fake utterances using 8 audio generation methods: WaveGlow~\cite{waveglow}, Multi--band MelGAN, (along with Full--band MelGAN)~\cite{mb-melgan}, MelGAN (along with MelGAN Large)~\cite{melgan}, HiFi--GAN~\cite{hifi-gan} and ParallelWaveGAN~\cite{parallel-wavegan} (with TTS pipeline).
Some of the methods are a modification of the others, e.g. 
TTS pipeline is based on the conformer~\cite{conformer} followed by a fine-tuned PWG; MelGAN Large differs from MelGAN by using a greater receptive field, whereas Full--band and Multi--band MelGANs use different approaches to compute one of their losses.

All samples were preprocessed using the WaveFake's recipe. The procedure consisted of resampling to 16kHz mono followed by trimming all the silences longer than 0.2s. The length of each sample was later normalized to about 4s by either trimming or padding (based on repeating the sample). Finally, we conducted an oversampling procedure to ensure the balance between the classes (bona fide vs. fake).

\begin{figure*}[h!]
\centering
    \begin{subfigure}[b]{0.33\textwidth}
       \includegraphics[width=1\linewidth]{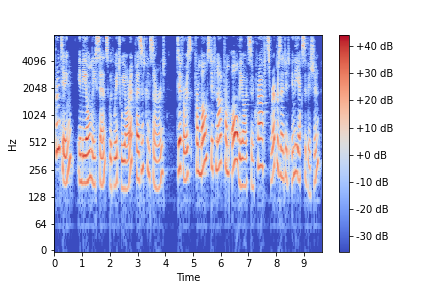}
       \caption{Spectrogram of the pristine sample.}
    \end{subfigure}

    \begin{subfigure}[b]{1\textwidth}
       \includegraphics[width=1\linewidth]{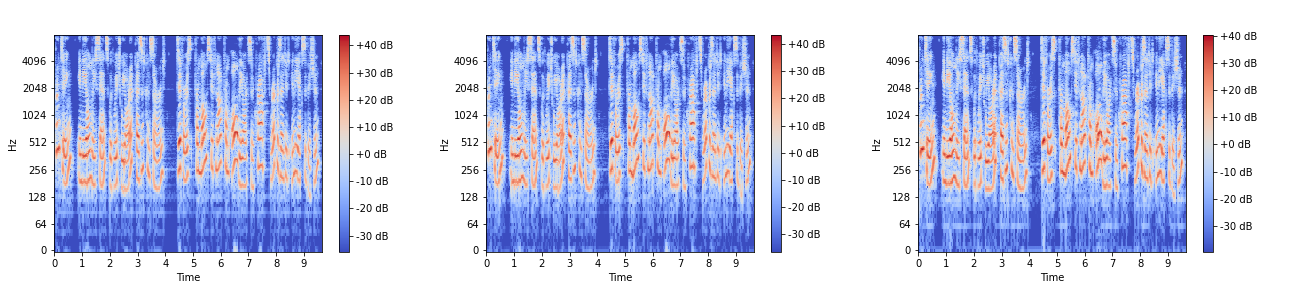}
       \caption{Spectrograms of MelGAN--Large, MelGAN and Parallel WaveGAN attacks. Choice was made basing on the results of Sect.~\ref{sec:limited_attacks}.}
    \end{subfigure}

    \begin{subfigure}[b]{1\textwidth}
       \includegraphics[width=1\linewidth]{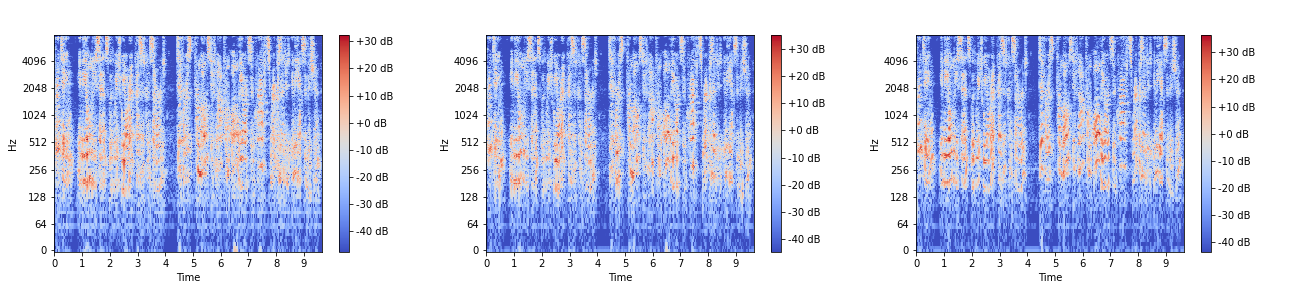}
       \caption{Differences between pristine spectrogram and respectively MelGAN--Large, MelGAN and Parallel WaveGAN modifications.}
    \end{subfigure}
    \caption{Spectrograms of the original sample from LJSpeech dataset along with corresponding WaveFake manipulations. We additionally report the absolute differences between spoofed and bona fide spectrograms.}
    \label{fig:our-activation-function}
\end{figure*}

\subsection{Procedure}

Each classifier was trained for 10 epochs --- each ended with a validation on a test set.
Checkpoint which scored the best validation results was selected for the final test on the eval set.
We follow the naming convention from the WaveFake paper and use train, test, and eval names for respective training, validation, and test sets. Each of the subsets --- training, test, and eval contained samples of every attack type, we used a dataset split of 70:15:15 ratio.
To make sure that the results are reproducible and generalize well, each architecture was trained and tested using 3 different random seeds. This is also applied to all of the remaining benchmarks.

The training was based on Adam optimizer~\cite{adam} with a batch size of 128 samples. We utilized the binary cross--entropy loss function. All the models were trained using a learning rate of $10^{-4}$ which was the original value reported in RawNet2 and LCNN papers.
Both RawNet2 and \model{} were additionally regularized using weight decay of value $10^{-4}$.

As mentioned, the models differed in the representation of the data they processed. While RawNet2 used a raw audio signal, LCNN and \model{} were based on spectrogram--based front--end, namely LFCC. 
They were created based on the parameters used by FakeAVCeleb authors~\cite{fakeavceleb} --- 25ms Hann windows with a 10ms window shift, 512 FFT points, and 80 LFCC features. This resulted in a two--dimensional feature of size $80 \times N$ ($N$ being the number of frames).

We based our comparison on two metrics --- Equal Error Rate (EER) and Area Under Curve (AUC).
Equal Error Rate (EER) is commonly used in biometry, DeepFake detection, and anti--spoofing tasks. In particular, we show EER as a percentage.
Area Under Curve (AUC) of Receiver operating characteristic (ROC) curve is a metric that reflects a statistical match well and is commonly used in binary classification tasks. Moreover, it takes into account all possible thresholds and shows their influence on the final results. 

\subsection{Results}

Table~\ref{tab:full_dataset} contains eval results of LCNN, RawNet2 and \model{} architectures trained on WaveFake dataset. 

\begin{table}[htb]
    \caption{Mean EER (\%) and mean AUC along with Std for the architectures tested on eval set of the standard benchmark. Values are averaged across three randomness seeds.}\label{tab:full_dataset} \centering
    \begin{tabular}{|c|c|c|}
      \hline
      \textbf{Model} & \textbf{EER (Std)} & \textbf{AUC (Std)} \\
      \hline
      LCNN & 0.1399 ($\pm$ 0.0325) & 99.9952 ($\pm$ 0.0031) \\ 
      \hline
      RawNet2& 4.5973 ($\pm$ 0.2742) & 99.1254 ($\pm$ 0.0670) \\
      \hline
      \model{} & 0.1549 ($\pm$ 0.0283) & 99.9941 ($\pm$ 0.0015) \\
      \hline
    \end{tabular}
\end{table}

LCNN provided the best results in terms of both metrics --- 0.1399 (EER) and 99.9952 (AUC).
\model{} was a close second scoring EER of 0.1549 and AUC of 99.9941. It is worth mentioning that the results of \model{} were more consistent (characterized by smaller standard deviation), which is one of the key factors in selecting DeepFake detection models.
RawNet2 showed the weakest performance scoring EER equal to 4.5973 and AUC of 99.1254. These results show that LFCC--based methods perform better than raw--signal architectures in the task of DeepFake detection.

\subsection{Time and \& complexity comparison}

Apart from evaluating the performance, we measured the complexity of the models. Table~\ref{tab:no_trainable_parameters} presents number of trainable parameters for each of the architectures. \model{} contains almost 2 times fewer parameters than LCNN and about 60 times less than RawNet2. This contributes to both faster inference times as well as lower memory requirements.

\begin{table}[htb]
\caption{The number of trainable parameters in each of the evaluated architectures.}\label{tab:no_trainable_parameters} \centering
\begin{tabular}{|c|c|}
  \hline
  \textbf{Model name} & \textbf{Trainable parameters} \\
  \hline
  \model{} & 277,963 \\
  \hline
  LCNN & 467,425 \\
  \hline
  RawNet2 & 17,620,385\\
  \hline
\end{tabular}
\end{table} 

In addition to calculating the number of the parameters, we investigated the inference times of the models on random samples.
We report results averaged across 1,000 inferences. To cover different use-cases, we conducted this benchmark on two types of devices --- GPU (NVIDIA Tesla P40) and CPU (2 GHz Quad--Core Intel Core i5). 
We report times on different batch sizes as it corresponds to the speed of validating the full material. The whole material can be analyzed with a single forward pass by splitting the utterance into subsequent audio chunks, which later can be used as batched input.

\begin{table}[tb]
    \caption{The comparison of inference times (milliseconds) using CPU and GPU in relation to the used batch size (BS). Greater batch sizes contribute to more significant differences between~\model{} and other models.}\label{tab:inference_times}
    \begin{center}
    \begin{tabular}{|c|c|c|c|}
    \hline
    \textbf{Model} & \multicolumn{3}{c|}{\textbf{CPU}} \\
    \cline{2-4} 
    \textbf{Name} & \textbf{\textit{BS 1}}& \textbf{\textit{BS 16}}& \textbf{\textit{BS 32}} \\
    \hline
    LCNN                        & 41.632  & 451.745  & 944.672  \\ \hline
    RawNet2                     & 151.913 & 1386.375 & 2637.200   \\ \hline
    SpecRNet                    & 27.358  & 370.843  & 706.300   \\ \hline
    
    \textbf{Model} & \multicolumn{3}{c|}{\textbf{GPU}} \\
    \cline{2-4} 
    \textbf{Name} & \textbf{\textit{BS 1}}& \textbf{\textit{BS 16}}& \textbf{\textit{BS 32}} \\
    \hline
    LCNN                        & 3.713  & 11.764 & 20.088 \\ \hline
    RawNet2                     & 11.934 & 43.317 & 56.787 \\ \hline
    SpecRNet                    & 3.669  & 7.1492  & 13.244 \\ \hline
    \end{tabular}
    \end{center}
\end{table}

Values presented in Table~\ref{tab:inference_times} (for both GPU and CPU cases) show that the larger batch size contributes to increasingly notable disproportion between inference times.
While the differences between LCNN and \model{} on a single element batch are negligible, they become more significant in the case of 16 and 32 samples. RawNet2 was characterized by the highest inference times --- up to about 6 times longer than those of \model{}. 
While RawNet2 operates on a raw waveform and requires no additional computations, both LCNN and \model{} are based on the LFCC front--end, which requires additional computations before using it as an input to the network. Calculation time was the same for both models and takes, on the GPU, respectively 0.9779ms, 7.2666ms, 13.6549ms for batch sizes of 1, 16 and 32, which still makes these solutions notably faster than RawNet2. The same principle applied to CPU calculations.
The overall results show that \model{} requires about 13ms to analyze a 2 minute material using GPU. Such efficiency proves that this model could be successfully used in the big data setting of analyzing materials uploaded to online multimedia platforms. Our DeepFake detection model may work on a processing buffer during the audio uploading, allowing nearly real-time verification of a sample, before making it accessible to the wider public.

\section{Bechmark --- limited attacks} \label{sec:limited_attacks}

The following section contains information about our benchmark exploring how various DeepFake manipulations (attacks) influence the performance of the final models. We ran each training omitting one of the available attacks i.e. training, test and evaluation procedures were conducted 8 times --- each time without one of the manipulations. The comparison includes results obtained on the full dataset.

Such limited attacks setting could give information about relations between particular attacks and the efficacy of the investigated methods. Presented results could be interpreted as follows: high EER values mean that a given attack was detected correctly in the baseline scenario (one could think of it as an 'easy to detect attack') and that its absence in the limited attack scenario decreases the performance.
Similarly --- lower EER suggests that given attack is difficult as its removal in a limited attacks scenario improves the results i.e. its presence in the full dataset had a negative impact on the yielded results. In such a case one may opt to train a dedicated methods that correctly (and with high efficacy) identifies this attack and when it is detected, the result of the specialized methods overrides the general DeepFake detection model.

This setting is similar to the one presented in~\cite{wavefake} and called out-of-distribution. While the training procedure is analogous (omitting one of the attacks), the two differ in evaluation procedure --- WaveFake's benchmark performs it additionally on the remaining attack. The main idea of that scenario was to evaluate generalization, whereas our benchmark focuses on assessing the difficulty of particular attacks for given models.

Our benchmark differs from the one described in Sect~\ref{sec:full_dataset_benchmark} in the attacks used in the training, test and evaluation procedures and the models that were evaluated. Setting concerned two spectrogram--based models: LCNN and \model{}, as they provided performance better than raw--audio--based RawNet2.

\begin{center}
\begin{table*}[]
\caption{EER values scored throughout limited attacks benchmark. Each column corresponds to the scenario of omitting particular generation method. We include full dataset results for reference.}\label{tab:drop_dataset_benchmark} \centering
\begin{tabular}{|c|c|c|c|c|c|}
\hline
\textbf{Model}    & \textbf{Full DS} & \textbf{HifiGAN}  & \textbf{FB--MelGAN} & \textbf{MelGAN--Large} & \textbf{PWG}     \\ \hline 
LCNN     & 0.1399 ($\pm$ 0.0325) & 0.0974 ($\pm$ 0.0079)  & 0.1014 ($\pm$ 0.0256) & 0.1483 ($\pm$ 0.0156)  & 0.1224 ($\pm$ 0.0227) \\ \hline
SpecRNet & 0.1549 ($\pm$ 0.0283) & 0.1367 ($\pm$ 0.0223)  & 0.1442 ($\pm$ 0.0655)  & 0.1565  ($\pm$ 0.0830)  & 0.1804 ($\pm$ 0.0382)  \\ \hline

\hline
\textbf{Model}    & \textbf{Full DS} & \textbf{MB--MelGAN} & \textbf{WaveGlow} & \textbf{MelGAN}   & \textbf{TTS} \\ \hline
LCNN     & 0.1399 ($\pm$ 0.0325) & 0.1234 ($\pm$ 0.0227) & 0.1188 ($\pm$ 0.0314) & 0.0871 ($\pm$ 0.0120) & 0.0942 ($\pm$ 0.0297)  \\ \hline
SpecRNet & 0.1549 ($\pm$ 0.0283) & 0.1459 ($\pm$ 0.0621) & 0.1676 ($\pm$ 0.0747)   & 0.1361  ($\pm$ 0.0301) & 0.1402 ($\pm$ 0.0826) \\ \hline

\end{tabular}

\end{table*}
\end{center}

Table~\ref{tab:drop_dataset_benchmark} contains results scored in the limited attacks benchmark. Values show differences between results in relation to the attacks in the dataset --- while some contribute to better results, others decrease them.
Performances of LCNN show that MelGAN was the most difficult for this classifier --- its absence improved EER from 0.1399 (full dataset) to 0.0871. On the other hand, the weakest EER of 0.1483 was scored when omitting MelGAN--Large. Interestingly this attack, as the only one, contributed to the metric increase. 
MelGAN was also the most difficult attack for \model{} --- it contributed to the decrease in EER from 0.1549 to 0.1361. Meanwhile, the absence of Parallel WaveGAN resulted in the highest EER of 0.1804, suggesting that these samples were, on average, detected best in the baseline scenario. 

AUC metric yielded comparable results for all of the evaluated scenarios --- they were greater than 99.99. The relative difference between the best and worst results were equal to 0.0042\% and 0.0045\% for respectively LCNN and \model{}. 
Absence of the attacks contributed to better results in cases of both LCNN and \model{}.
LCNN scored the AUC of 99.9952 on full dataset, whereas absence of Hifi--GAN resulted in the highest AUC of 99.9994. Similarly baseline AUC of \model{} was equal to 99.9941, whereas lack of MelGAN provided AUC of 99.9987. The full table was omitted due to manuscript's space constraints.

There was a disproportion between the results achieved in cases of omitting MelGAN and MelGAN--Large (alternative of MelGAN with a bigger receptive field). We conclude that the larger architecture of MelGAN--Large contributes to a more significant number of produced artifacts i.e., features used to differentiate between fake and bona fide samples.
The performance of our model was stable and close to LCNN's. In both cases, there existed no scenario in which a classifier learns only to differentiate one attack and therefore neglects other manipulations --- all of them were taken into account.
It is also worth mentioning that the lower Std of EER characterized \model{} for the results for the different attack--omitting scenarios --- 0.0149, whereas LCNN scored 0.0189. This indicates that our architecture is characterized by better generalization properties and should provide stable performance in the case of other new attack methods.

\section{Benchmark --- short utterances}\label{sec:short_utterances}

The difficulty of determining the validity of the utterance is influenced by many factors. They can be divided into technical like the quality of manipulation method (e.g. phoneme mapping or vocoding quality) and environmental like interference such as traffic noise or distortion in phone calls. Duration of the utterance is another crucial factor --- longer samples containing several words (or even sentences) tend to show more imperfections like inconsistencies or artificiality in prosody, pronunciation, and sudden frequency or tone change. 
In this benchmark, we aimed to evaluate the efficacy of the networks when trained on shorter sequences. Such a setting reflects a scenario when an adversary, instead of generating a whole utterance, replaces keywords of the sentence with generated words. This way, the utterance maintains the naturalness of the original, yet addresses different entity or action e.g. politician, instead of criticizing the opponent, denounces their own ally.

The discussed benchmark and the one introduced in Sect.~\ref{sec:full_dataset_benchmark} differ in the duration of samples --- utterances from all subsets were trimmed to 1s (instead of default 4s). Note that shorter samples are more difficult to identify as a DeepFake, yet 1s is enough to spot a keyphrase like a password during speaker identification or a name, or an important phrase during fake news generation. In addition, we ensured that all samples contained speech, it is achieved by trimming silences in earlier preprocessing stages.

\begin{table}[tb]
\caption{EER and AUC values scored by the models trained and tested using short utterances dataset (1s instead of 4s).}\label{tab:1_second_dataset} \centering
\begin{tabular}{|c|c|c|}
\hline
\textbf{Model name} & \textbf{EER (Std)} & \textbf{AUC (Std)}  \\
\hline
LCNN         & 0.9955 ($\pm$ 0.3744) & 99.9486 ($\pm$ 0.0324) \\ \hline
RawNet2      & 14.4499 ($\pm$ 0.8415) & 93.4013 ($\pm$ 0.7183) \\ \hline
\model{}     & 1.1781 ($\pm$ 0.0952) & 99.9322 ($\pm$ 0.0324)    \\ \hline
\end{tabular}
\end{table}

Table~\ref{tab:1_second_dataset} shows that all the architectures performed worse on shorter utterances. LCNN provided results most similar to the baseline benchmark by scoring EER and AUC of respectively 0.9955 and 99.9486. The differences in the case of \model{} were more significant --- 1.1781 and 99.9322. RawNet2 was the least stable solution having an EER of 14.4499 and AUC of 93.4013. 
The main reason behind the decrease in the performance was that classifiers operated on 4 times smaller data volume. This led to a smaller number of potential artifacts that differentiate between bona fide and generated samples. Nonetheless, spectrogram--based methods were robust enough to provide satisfying results in such a sophisticated scenario. This means that these models can be successfully used in the cases of the manipulations based on e.g. targeted substitution of keywords in order to produce fake news or impersonate the speaker during verification.

\section{Benchmark --- data scarcity setting}\label{sect:scarcity}

New DeepFake generation methods are constantly being introduced. The following benchmark covered training using a significantly smaller amount of the data. Such a scenario reflects a situation when some new DeepFake manipulations are published, but there is no access to their generation procedure or in the case of in--house verification using consumer--grade PC. In such case, detection methods must base on the gathered samples --- their quantity would be significantly smaller than in a standard case when the dataset contains samples additionally generated by scientists.

This benchmark was based on the one described in Sect.~\ref{sec:full_dataset_benchmark}. In the process the amount of the data used in training was reduced --- 10\% of the original train and test subsets were used. Moreover, in order to prevent overfitting on a smaller dataset, we reduced the number of epochs from the original 10 to 4. Note that the testing on the eval dataset was performed using a full number of samples as the attacker is not restricted when using their own attacks, hence we decided to use the full evaluation as more representative.

\begin{table}[tb]
\caption{Test results obtained by the architectures trained during data scarcity setting on 10\% of the training data.}\label{tab:10_percent_dataset} \centering
\begin{tabular}{|c|c|c|}
  \hline
  \textbf{Model} & \textbf{EER (Std)} & \textbf{AUC (Std)} \\
  \hline
  LCNN     & 0.6305 ($\pm$ 0.1262) &  99.9599 ($\pm$ 0.0148) \\  \hline
  RawNet2  & 24.0821 ($\pm$ 1.9119) &  84.3158 ($\pm$ 2.0555) \\ \hline
  \model{} & 0.7997 ($\pm$ 0.1918) & 99.9390 ($\pm$ 0.0290) \\   \hline
\end{tabular}
\end{table}

Table~\ref{tab:10_percent_dataset} shows a performance decrease across all solutions in relation to the ones scored on the full dataset.
Both LCNN and \model{} provided acceptable performance despite a 10 times smaller volume of training data --- respectively, EER of 0.6305 and 0.7997 along with AUC of 99.9599 and 99.9390. On the other hand, RawNet2 provided weaker performance --- EER of 24.0821 and AUC of 84.3158. One of the reasons behind that was the network size --- this architecture contains over 17 million parameters which requires a more significant number of training samples to generalize well.

Nonetheless, results of LCNN and \model{} indicate their robustness against the small quantity of the data and show that they are characterized by good generalization abilities --- only 10\% of the used samples were sufficient to capture underlying features used to distinguish bona fide and generated utterances. This information also suggests that new datasets for DeepFake detection should focus more on generation methods' number (and diversity) than a large number of samples.

Benchmarks from Sect.~\ref{sect:scarcity} and ~\ref{sec:short_utterances} operating on a reduced data volume (10\% and 25\% of total hours, respectively) differ in information they contain. RawNet2 scored better EER for shorter utterances than the data scarcity. This might be indicated by the number of model parameters --- data reduced by 90\% did not contain enough samples for it to generalize well. 
On the other hand, the data scarcity setting provided better EER results for LCNN and \model{}, however, the differences were not as significant as for RawNet2. We conclude that the number of samples in the dataset is sufficient for them to generalize. On the other hand, short utterances contain on average 75\% fewer artifacts, which decreases the performance.

\section{Conclusions and Future Work}

In this work, we addressed the improved time performance in the task of audio DeepFake detection. 
Our main contribution is providing the novel neural network architecture --- \model{}, that despite a significant decrease in computational requirements, is still capable of achieving results comparable to state--of--the--art models. Our method has a wide variety of applications that include processing a high volume of audio--visual materials in online multimedia platforms, as well as being the solution available to use by average citizens to verify materials on their own using their personal devices.

\model{} is a novel architecture inspired by RawNet2 and based on spectrogram representation of audio. Our model provides up to about 40\% improvement in inference speed in relation to one of the fastest DeepFake detection methods --- LCNN model while achieving comparable detection results. In the basic benchmark tests, it was only $0.001\%$ worse (in terms of AUC) than LCNN, and the difference on reduced datasets (both, the trimmed samples and the reduced number) did not exceed $0.02\%$. 
Our comparison included \model{} with RawNet2 and LCNN networks using a benchmark conducted on the WaveFake audio DeepFake dataset.

In addition, we verified the robustness of the aforementioned architectures against three novel scenarios --- limited attacks, utterances of shorter length, and a small quantity of the training data. 
The first provided more profound insight into the detection task's difficulty (performance) in relation to every DeepFake generation method. 
The two other benchmarks were novel ways to evaluate the model in more sophisticated scenarios --- in the case of keywords replacement (shorter utterances) and the scenario of smaller datasets. Those tests could be of significant importance in the case of targeted attacks (aimed for a specific phrase or a password), in the case of attacks that lack the data for a standard training regime (e.g., for new attacks), and for users that want to verify the samples by themselves, thus are constrained by the limitations of consumer--grade PCs. Moreover, such detection procedure may be used in the content providing platforms both in the case of the fast prescreening method, when the volume of data is too big for full verification, or when the ratio between the training and eval sets is imbalanced, like in the case of Sect.~\ref{sect:scarcity}.
Both additional benchmarks proved that despite a drop in performance resulting from the more challenging scenarios; our architecture still provides decent performance.

Future work in this field should focus on improving the model's performance and evaluating its generalization capabilities.  Additionally, one may focus on specifically targeting speaker recognition systems utilizing passphrases that are interwoven into full sentences (cf.~\cite{9339931,LASKAR202029,LASKAR2021115281}). 
To further enhance the network's results, one may utilize another front--end features e.g. mel--frequency cepstral coefficients (MFCC) or a combination of such features, which was proved to provide a better stabilization along with generalization~\cite{attack-agnostic-dataset}.
Generalization can also be assessed on other existing audio DeepFake datasets such as ASVspoof2021 and FakeAVCeleb or a combination of them --- Attack Agnostic Dataset~\cite{attack-agnostic-dataset}. This approach is easily scalable and therefore can be used with new datasets. Thanks to its use of different attacks in various dataset folds, Attack Agnostic Dataset provides a good overview of the potential model's performance in the face of new DeepFake generation algorithms. 

\bibliographystyle{IEEEtran}
\bibliography{bibliography}

\end{document}